# Bottleneck Analysis of Dynamic Graph Neural Network Inference on CPU and GPU

Hanqiu Chen[1], Yahya Alhinai[1§], Yihan Jiang[1§], Eunjee Na[2§], Cong Hao[1]
{hchen799,yhinai3,yjiang400}@gatech.edu, jasmin7907@kaist.ac.kr, callie.hao@gatech.edu

[1]*School of Electrical and Computer Engineering, Georgia Institute of Technology*
[2]*Korea Advanced Institute of Science & Technology*

## Abstract

*Dynamic graph neural network (DGNN) is becoming increasingly popular because of its widespread use in capturing the dynamic features in the real world. A variety of dynamic graph neural networks designed from algorithmic perspectives have succeeded in incorporating temporal information into graph processing. Despite the promising algorithmic performance, deploying DGNNs on hardware presents additional challenges due to the model complexity, diversity, and the nature of the time-dependency. Meanwhile, the differences between DGNNs and static graph neural networks make hardware-related optimizations for static graph neural networks unsuitable for DGNNs. In this paper, we select eight prevailing DGNNs with different characteristics and profile them on both CPU and GPU. The profiling results are summarized and analyzed, providing in-depth insights into the bottlenecks of DGNNs on hardware and identifying potential optimization opportunities for future DGNN acceleration. Followed by a comprehensive survey, we provide a detailed analysis of DGNN performance bottlenecks on hardware, including temporal data dependency, workload imbalance, data movement, and GPU warm-up. We suggest several optimizations from both software and hardware perspectives. This paper is the first to provide an in-depth analysis of the hardware performance of DGNN*. Code is available at https://github.com/sharc-lab/DGNN_analysis.*

## 1. Introduction

Deep neural networks (DNNs) have made tremendous breakthroughs in various domains such as speech [1–3], image [4–7], and natural language processing [8–10]. While DNNs can effectively capture the hidden pattern in Euclidean data, they do not perform well in processing non-Euclidean data presented in graph format, such as social networks, recommendation systems, and knowledge graphs. Facing the challenges of DNNs, researchers have an increasing interest in the graph data processing. Graph Neural Network (GNN) is a powerful tool for processing graphs. GNNs have shown the ability to capture the expressive power of graphs in many research areas. These areas include social networks [11], physical systems [12] and new drug discovery [13]. Fig. 1(a-b) demonstrates how a social network can be represented as a graph, and how GNNs can be applied to the graph. This is called *static* GNN, in which both the graph and the model parameters do not change.

Although GNNs have a strong representation power on static graphs, in real-life, most of the graphs are changing with time. For instance, as shown in Fig. 1 (c), social networks are constantly growing as more people are joining, and the edge weights are also changing as the relationships between people evolve. In this case, the graph topology changes over time, with nodes and edges appearing and disappearing as time progresses. To better represent complex graph structures changing over time, dynamic graph neural network (DGNNs) is a promising solution. Fig. 1(c) illustrates an example of a dynamically changing graph, and Fig. 1(d) is an example of a *dynamic* GNN, which uses a graph structure encoder to capture the spatial information and a time encoder to encode the temporal information.

According to a recent survey [14], DGNNs can be divided into two categories according to the graph representation temporal granularity: (i) discrete time dynamic graph neural network (DTDG) and (ii) continuous time dynamic graph neural network (CTDG). DTDG captures the status of a changing graph at different time steps and uses the information from multiple snapshots of the dynamic graph to gain a deeper understanding of the graph features. The algorithm takes into account the temporal information to better understand the features of the graph. CTDG is an event-based neural network that can update node and edge embeddings at any time when an edge between two nodes appears. This update method is more realistic because it is closer to real-world scenarios.

Although DGNNs have achieved great success from an algorithmic perspective, their hardware performance is extremely poor due to the dire lack of hardware optimization. Because of hardware bottlenecks, the lack of optimization results in high latency and a low degree of parallelism.

Motivated by the software-hardware performance gap of DGNNs, this paper seeks to provide a quantitative analysis of eight representative models with different characteristics.

*Hanqiu Chen profiled EvolveGCN and TGAT models and led paper writing. Yahya Alhinai profiled JODIE, TGN, and MolDGNN models. Yihan Jiang profiled DyRep and LDG models. Eunjee profiled ASTGNN model and prepared the codebase; work done during her intern at Georgia Tech. § Equal contribution.

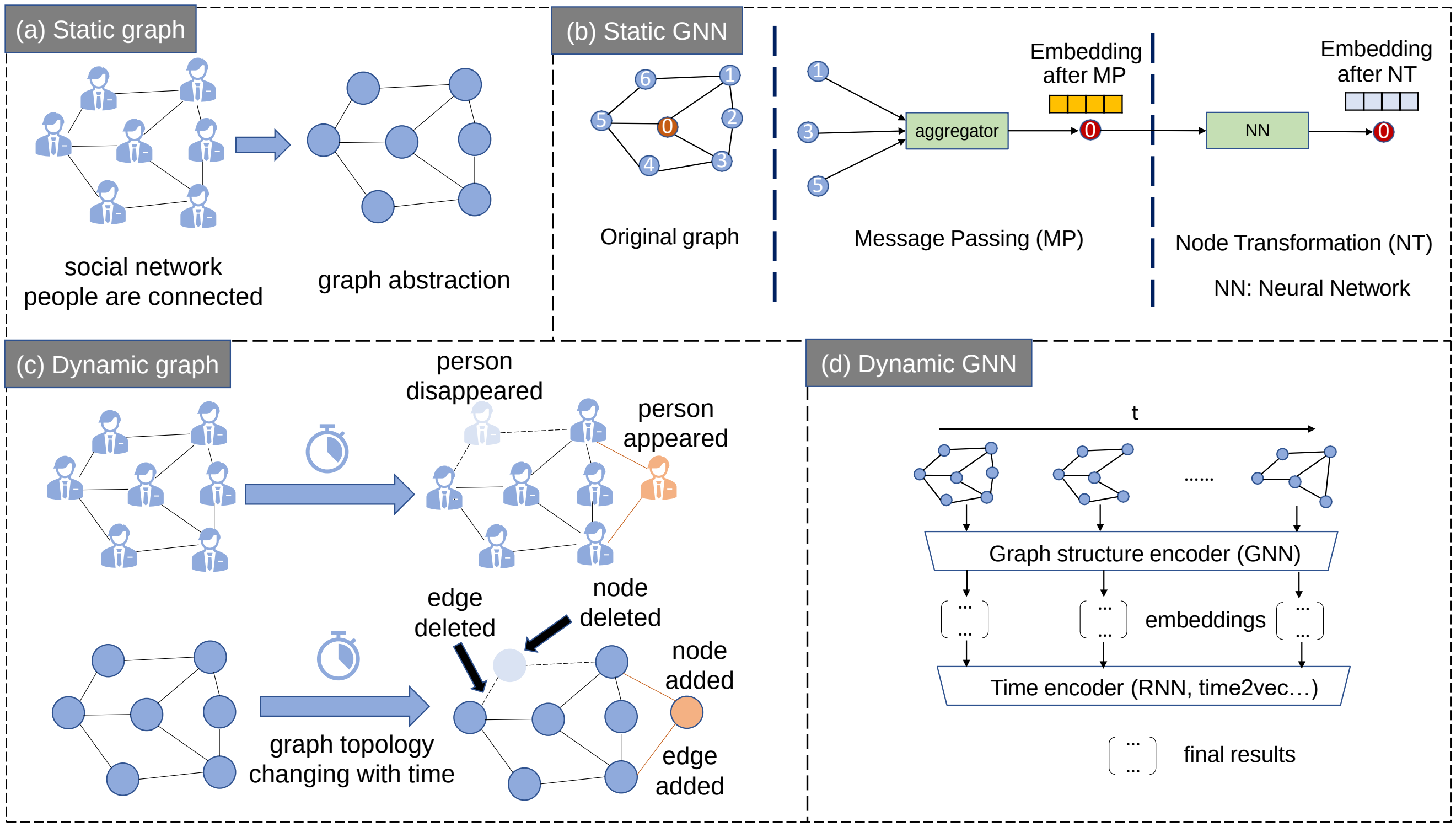

**Figure 1: Introduction to graph neural networks (GNNs) and dynamic graph neural networks (DGNNs). (a) Static graph modelling of a social network. (b) An example structure of a static GNN with message passing (MP) and node transformation (NT). (c) Dynamic graph modelling of a social network varying with time. (d) A high-level architecture of a DGNN with a graph structure encoder to capture spatial information and a time encoder to capture temporal information.**

By profiling these models on both CPU and GPU, we aim to provide root causes behind the hardware bottlenecks during inference, as well as further acceleration optimizations. The models analyzed are: JODIE [15], EvolveGCN [16], TGAT [17], MolDGNN [18], ASTGNN [19], DyRep [20], TGN [21], and LDG [22].

To date, little research has been conducted on the hardware performance of DGNNs. Thus, this paper provides the first comprehensive DGNN performance analysis on hardware with detailed bottleneck analysis. Our contributions are summarized as follows:

1) **Comprehensive review.** We first provide an overview of the background of GNNs and DGNNs. We then introduce eight representative DGNNs with their high-level dataflow and data dependency graphs to explain their computational behavior, which largely impact the hardware performance.
2) **Hardware profiling.** We use PyTorch Profiler [23] and NVIDIA Nsight Systems [24] to profile eight DGNN models on both CPU (6226R) and GPU (A6000). PyTorch Profiler is used to investigate memory usage and execution time of different modules in different DGNNs; Nvidia Nsight Systems is used for detailed CUDA and kernel operations breakdown and analysis.
3) **DGNN hardware bottleneck analysis.** We showcase and analyze the profiling results of DGNN inference on both CPU and GPU. We identify four common hardware bottlenecks that will degrade DGNN hardware performance: ❶ Temporal data dependencies between different snapshots, node embedding, and edge embedding lead to limited parallelism on hardware, which causes low GPU utilization and degrades GPU acceleration rate. ❷ The time-evolving graph topology of DGNNs causes irregular memory accesses and a high graph sampling overhead. This leads to a workload imbalance between CPU and GPU, which makes CPU or GPU computation kernels idle and limits high hardware utilization. ❸ The time-evolving graph topology, node and edge embedding cause frequent data exchange between CPU and GPU. ❹ Before inference starts on GPU, GPU needs to warm up because of model initialization and memory allocation on GPU.
4) **Proposal of promising optimizations.** We propose several optimizations that can help overcome the DGNN hardware bottlenecks uncovered in this work. We classify them into three categories according to their application scope: ❶ **software optimizations** including reducing graph preprocessing overhead and implementing smarter framework; ❷ **hardware optimizations** including improving parallelism on GPU and reducing data transferring overhead between CPU-GPU; ❸ **hardware/software co-design** including extending current static GNN hardware optimizations to DGNNs and designing hardware-specific accelerators to improve DGNNs hardware efficiency.

TABLE 1: Summary of the DGNNs profiled in this work. Column 2 is the type of a DGNN; column 3 $\sim$ 6 show which feature is evolving with time; column 7 is the time encoding method used in DGNN; column 8 gives examples of the tasks the DGNN can be applied to.

| DGNN | type | node feature | edge feature | graph topology | weights | time encoding | tasks |
|---|---|---|---|---|---|---|---|
| JODIE [15] | continuous | ✓ | | ✓ | | RNN | future interaction prediction<br>state change prediction |
| TGN [21] | continuous | ✓ | | ✓ | | time embedding | future edge prediction |
| EvolveGCN [16] | discrete | ✓ | | ✓ | ✓ | RNN | link prediction<br>node classification<br>edge classification |
| TGAT [17] | continuous | ✓ | ✓ | ✓ | | time embedding | link prediction<br>link classification |
| ASTGNN [19] | discrete | ✓ | | ✓ | | self-attention | traffic flow prediction |
| DyRep [20] | continuous | ✓ | ✓ | ✓ | | RNN | dynamic link prediction<br>time prediction |
| LDG [22] | continuous | ✓ | ✓ | ✓ | ✓ | RNN<br>self-attention | dynamic link prediction |
| MolDGNN [18] | discrete | | ✓ | ✓ | ✓ | RNN | adjacency matrix prediction |

The rest of the paper is organized as follows. Section 2 clarifies the current development of hardware accelerators for GNNs and outlines the motivation of this paper. Section 3 provides detailed information about the data flow and data dependency graph for eight DGNN models for profiling. Section 4 presents profiling results with bottleneck analysis. Section 5 proposes various optimizations in terms of software, hardware, or a combination of both software and hardware (software/hardware co-design) in order to address the bottlenecks associated with the graph topology of DGNNs. Section 6 summarizes the work presented in this paper, points out the limitations of this research, and discusses future directions.

## 2. Motivations

Aiming to gain orders-of-magnitude performance improvement, various GNN accelerators have been proposed to handle the computation- and communication-intensive GNN challenges targeting different hardware platforms including CPUs, GPUs, ASICs, FPGAs, and heterogeneous architecture. To speed up the sparse matrix multiplication (SpMM) for embedding transformation, which is widely used in GNNs, AWB-GCN [25], I-GCN [26], and BoostGCN [27] are proposed. Addressing the irregular memory access in the message processing, the following methods have been proposed: GraphACT [28], HyGCN [29], and VersaGNN [30]. These methods all exploit data locality or partition the graph on the CPU in order to address the issue.

Despite the fact that GNN accelerators have gained state-of-the-art performance in hardware in recent years, they are only applicable to static graphs. Some DGNN hardware accelerators [31, 32] have been proposed recently, however, there still remain some challenges on DGNN hardware deployment introduced by dynamic graphs that have not yet been addressed yet. We believe the hardware analysis of DGNN models we presented in this paper is meaningful and significant for two reasons:

- Most of the existing DGNNs are designed from the algorithmic perspective. They only focus on harnessing the network structures' temporal encoding methods to improve the performance of DGNNs on specific tasks. The lack of attention to hardware efficiency induces huge computational overheads, low hardware parallelism, and high energy consumption. Identifying the hardware bottlenecks of existing DGNNs helps researchers avoid algorithm inefficiency and potentially leads to the design of hardware-compatible DGNNs in the future.
- Hardware optimizations in current GNN accelerators primarily focus on matrix multiplications, and achieve significant hardware performance improvement when processing static graphs. However, dynamic graph processing needs to take temporal data dependencies into consideration, which brings new challenges. The reason is that previous optimizations for static graph processing do not behave well when facing temporal data dependencies and we need to design new parallel computing schemes to overcome them.

## 3. Background

**Graph Neural Networks (GNNs)** are neural networks capable of leveraging the structure and properties of graphs in graph-based tasks. GNNs take graphs, endowed with node and edge features, as inputs and performs the computation based on their features and the graph structures. Prevailing GNN models follow the message passing mechanism, including two major stages: message passing and node transformation. In each layer of the GNN, these two steps will be applied to the input graph in turn. Fig. 1 (b) demonstrates a message passing procedure for a single node 0 at one of the layers. In the message passing stage, node 0 aggregates features from adjacent nodes by using a self-defined or learned aggregator, such as mean, average, or max. In the node transformation stage, the feature of node 0 is updated by a linear layer or a multi-layer perceptron (MLP). There are many variations of GNNs; a more detailed introduction can be found in recent surveys [33, 34].

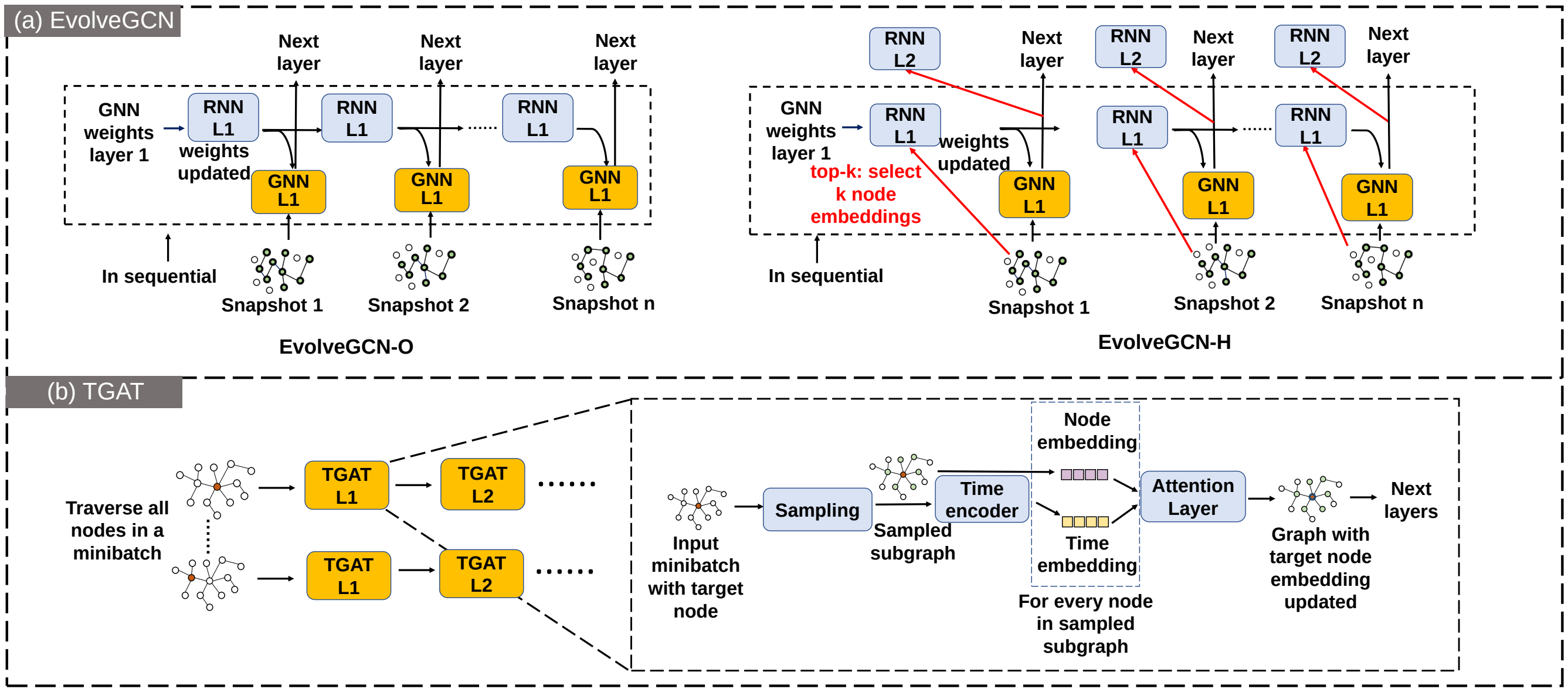

**Figure 2: The computation flow of (a) EvolveGCN and (b) TGAT.**

**Dynamic Graph Neural Networks (DGNNs)** are GNNs used to process the temporal information in dynamic graph-structured data, as shown in Fig. 1 (c) and (d). What separates DGNNs from GNNs is the innovative time encoder, such as Recurrent Neural Networks (RNNs) and Time2Vec [35], which encode time-series features of graphs varying with time.

DGNNs are specialized with various features, such as different graph structures and methods of time encoding, to get the best performance in individual real-world applications. In this section, we present a brief description of the 8 representative DGNNs with graphs of dataflow dependencies and a summarizing table 1 of their key features. We include the type of DGNN, which part is changing with time, the time encoding method and tasks that can be applied to in this table.

## 3.1. EvolveGCN

EvolveGCN [16] is a discrete time dynamic GNN. It can be applied to link prediction and node/edge classification tasks in social networks, where the graph topology and embedding are changing over time.

EvolveGCN takes a snapshot of the whole graph at each time step to process temporal subgraphs using a graph convolution network (GCN), and meanwhile adopts a recurrent neural network (RNN), e.g., gated recurrent unit (GRU) [36] and long short-term memory (LSTM) [37], to regulate the network parameters. There are two versions of EvolveGCN: EvolveGCN-H and EvolveGCN-O, and their architectures are shown in Fig. 2 (a). In each layer, the RNN and GNN are executed sequentially because GNN needs the weights renewed by RNN.

The difference between -H version and -O version is that, in -O version, the input of RNN is GNN weights while in -H version, both the GNN weights and node embedding are the input of RNN. Since the dimension of the node embedding matrix is inconsistent with that of the GCN weight matrix, an additional module (top-k) for node embedding sampling is needed to ensure the dimension consistency.

## 3.2. TGAT

Temporal Graph Attention Network (TGAT) is relevant for applications that predict the future behavior of nodes in a temporal graph and detect changes in node behavior over time. Examples of such applications include social networks, transportation networks, and biological networks.

TGAT is inspired by, first, the self-attention mechanism to aggregate neighborhood nodes' information and, second, classical Bochner's theorem of harmonic analysis to encode the temporal information. TGAT generates the time embedding based on two assumptions: critical temporal information is revealed in the relative time value and this value will be concatenated with the node embedding as the input of the attention layer to renew the target nodes' embedding. Detailed computation dataflow of TGAT is shown in Fig. 2 (b). TGAT traverses all of the nodes in the graph sequentially to update the node embeddings in the mini-batch. In each TGAT layer, the sampling, time encoding, and the attention layer are sequential.

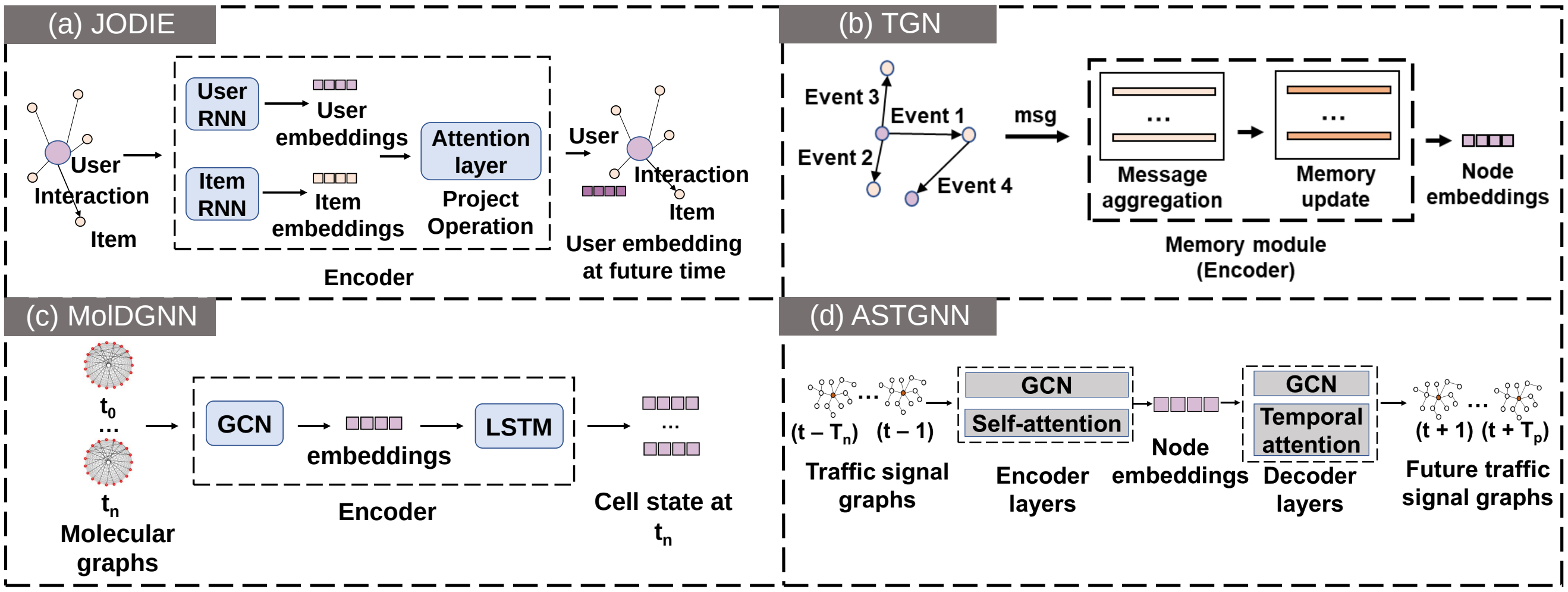

**Figure 3: The computation flow of (a) JODIE (b) TGN (c) MolDGNN (d) ASTGNN.**

## 3.3. JODIE

Predicting Dynamic Embedding Trajectory in Temporal Interaction Networks (JODIE) [15] is a continuous time dynamic graph neural network. JODIE is used to predict how a user's interests change over time, based on their past interactions within a network. It can be used to model the dynamics of social networks and to understand how information propagates through them. The model is designed such that the future dynamics of a system are represented as a low-dimensional embedding in a high-dimensional space. In particular, it uses a user RNN and an item RNN to predict the user and item embeddings, respectively, in the future.

To speed up the training process, JODIE's paper proposes a batching algorithm called t-batch. The algorithm creates inter-dependencies between nodes to separate them into batches to be processed in parallel. It is reported that t-batch speeds up JODIE by 9.2× [15]. In our profiling experiment, we implement t-batch to work on the inference of JODIE (interaction and state change prediction). This allows for a significant improvement in GPU utilization and overall performance.

A high-level overview of JODIE dataflow is shown in Fig. 3 (a). The user RNN and item RNN for embedding update are mutually-recursive. The embedding projection operator (attention layer) predicts the future embedding trajectory of the user. We also show the data movement details between CPU and GPU of JODIE in Fig. 5 (a). First, the t-batch is created on CPU. Then, it is transferred to GPU, where each node executes the embedding projection and prediction. Finally, the batch of nodes' projected embeddings is sent back to CPU, and the next t-batch is executed.

## 3.4. TGN

Temporal Graph Networks (TGN) [21] is proposed to learn about dynamic graphs by generalizing the Graph Neural Network (GNN) framework to the case of temporal graphs. This model could be applied in the fields of social sciences, recommender systems, and biological interaction networks in order to better understand complex networks and the relationships between the nodes therein.

TGN first samples a batch of node-to-node interactions according to the events that happened at a certain time step and then transfers this batch from CPU to GPU for the following computation. On GPU, it uses a memory module, including message aggregation and memory update, to capture the long-term temporal dependencies to generate the node embedding at any time. Node embeddings at any time $t$ zare produced by using the temporal graph and the node's memory. Detailed dataflow of TGN is shown in Fig. 3 (b).

There is a frequent memory exchange between CPU and GPU for TGN during the inference. We show this complex memory exchange using Fig. 5 (b).

## 3.5. MolDGNN

Molecular Dynamic Graph Neural Network (MolDGNN) [18] is a discrete time dynamic graph neural network for learning the dynamics of molecules. MolDGNN is relevant for applications in computational chemistry and physics, where the accurate prediction of molecular dynamics is essential. In particular, this model can be used to predict changes in molecular structure over time, which can provide deep insights into their functionality.

The key idea is to encode the molecules as graphs, where the nodes correspond to the atoms and the edges correspond to the chemical bonds between them. The dynamics of the molecule are then modeled as a sequence of graphs with different topologies, where each state corresponds to a particular configuration of the molecule. The sequence of molecular graphs are passed into the GCN-LSTM encoder. The GCN component models the topological structure of each graph snapshot, and the LSTM component models the temporal structure of the entire sequence. The GNN is used to learn the spatial relationship between graphs in different time steps, and the LSTM is used to capture the dynamics of the molecule's atoms. Detailed computation flow is shown in Fig. 3 (c). The communication between CPU and GPU is a major bottleneck of the model. The large data moved (adjacency matrix) from CPU to GPU and vice versa incur significant latency, as shown in Fig. 5 (c). Consequently, the system cannot take full advantage of the GPU's computational power.

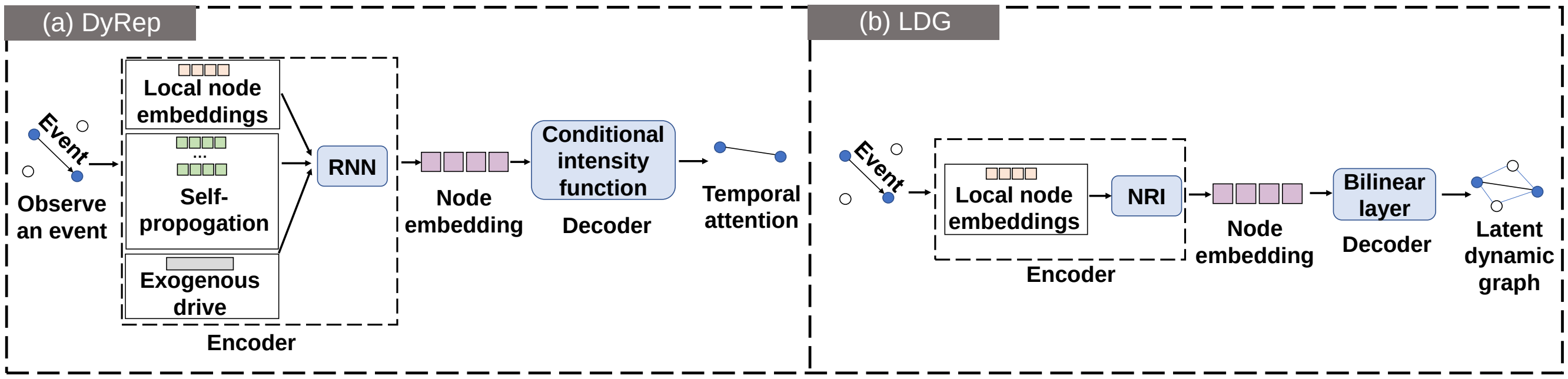

**Figure 4: The computation flow of (a) DyRep and (b) LDG.**

### 3.6. ASTGNN

Attention-based Spatial-Temporal Graph Neural Network (ASTGNN) [19] is a discrete time dynamic graph neural network for traffic forecasting. The model is designed to capture the dynamics of traffic. It can be used to predict traffic conditions in a given area and to understand how traffic information evolves in a road network.

ASTGNN has an encoder-decoder structure with multiple layers of alternating stacked blocks, each modeling temporal and spatial dynamics. Each encoder layer is built up with self-attention block and spatial dynamic GCN block; each decoder layer includes two temporal attention block and a spatial dynamic GCN block. The encoder maps a traffic signal matrix sequence to intermediate representations, and the decoder generates the future traffic signals by taking intermediate representation as input. The dataflow of ASTGNN is shown in Fig. 3 (d).

### 3.7. DyRep

Representation learning framework for dynamic graphs (DyRep) [20] is a continuous time dynamic graph neural network. It is designed to process temporal information in social network communications, biological networks, and infrastructure networks.

DyRep models interleaved evolution of short-term and long-term processes with temporal point patterns. The occurrence of an event is dependent on the most recent state of the graph. The RNN in the encoder will combine the message from local node embeddings, self-propagation, and exogenous drive together to update the node embedding. After obtaining the renewed node embedding, a conditional intensity function is used as a decoder to model the occurrence of an event between nodes at a specific time. Finally, temporal attention weights are learned from the decoder outputs. Fig. 4 (a) shows the whole process.

### 3.8. LDG

Latent Dynamic Graph (LDG) [22] is based on the DyRep [20] and integrated with NRI [38] to learn temporal attention between nodes by observing the dynamics of events. The application of this model is similar to DyRep, including modelling the evolution of social networks and tracking the spread of diseases in biological networks.

DyRep and LDG have the same node embedding phase. The major difference between them is the encoding phase. LDG chooses to use a sequence of learnable functions from NRI (Neural Relational Inference [38]) to do the node and edge mapping and generate the node and edge embedding. The decoder is a learnable bilinear layer used to update the generated node embedding. The computation dataflow of LDG is shown in Fig. 4 (b). The encoder takes node embedding at the previous time step as input and outputs the edge embedding between the nodes passed in. The bilinear decoder provides a richer interaction between embeddings of different nodes.

## 4. Experiments and Bottleneck Analysis

This section presents the profiling results of the eight DGNN models in the aspects of inference time comparison between CPU and GPU, GPU usage, memory usage, inference breakdown, and warm-up time. The bottlenecks revealed from the profiling results can be classified into four major categories: temporal data dependency bottleneck (Sec. 4.1) , workload imbalance bottleneck (Sec. 4.2), data movement bottleneck (Sec. 4.3), and GPU warm-up bottleneck (Sec. 4.4). For better understanding, a short introduction of each bottleneck will be provided before the detailed profiling experiment analysis.

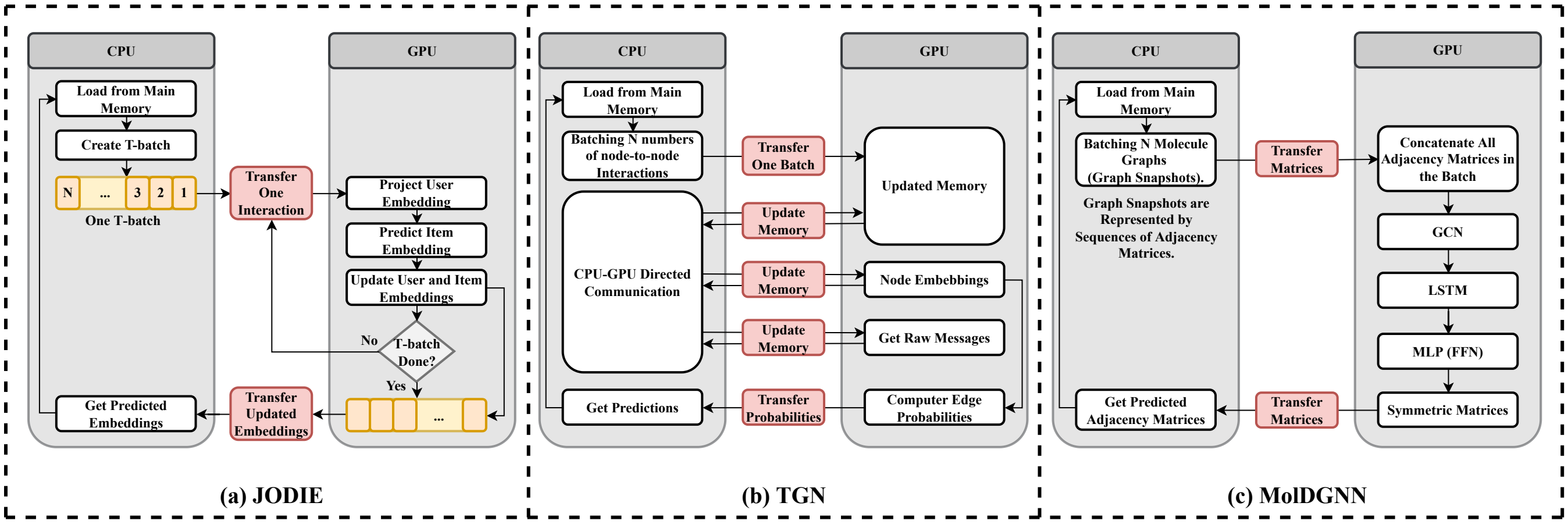

**Figure 5: High-level overview of CPU-GPU communication of (a) JODIE (b) TGN (c) MolDGNN.**

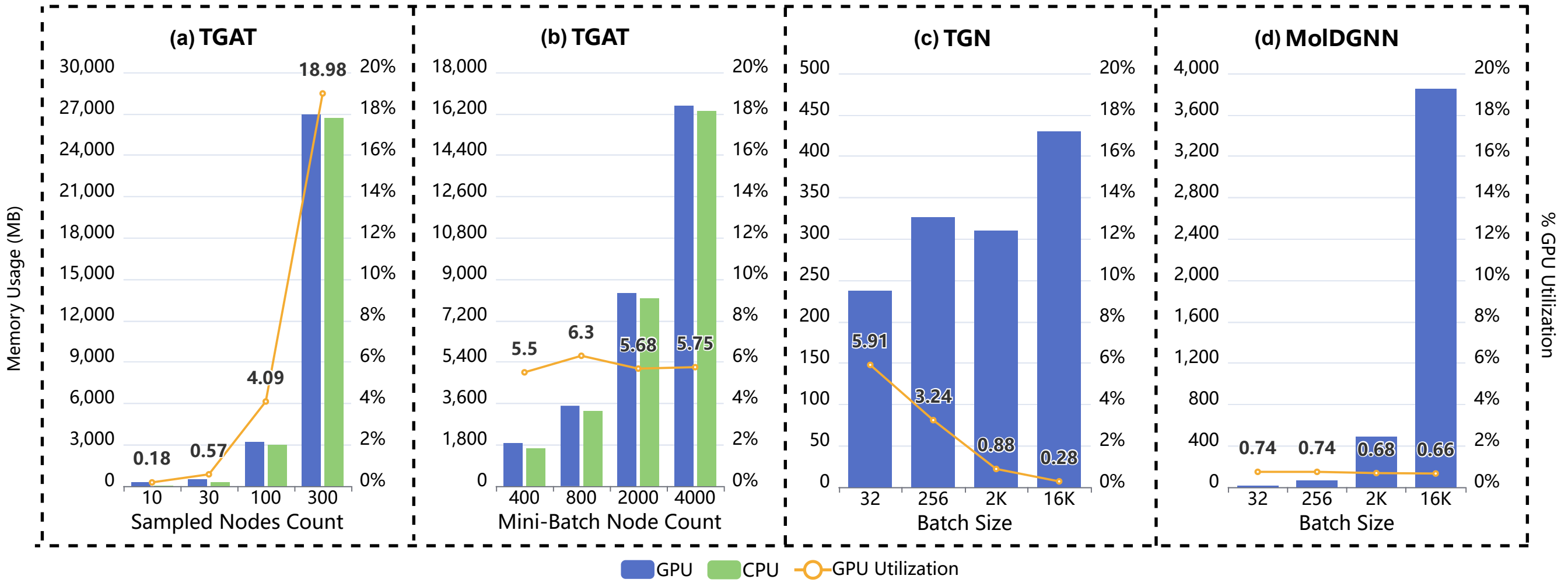

**Figure 6: Memory usage and GPU utilization of different models. (a) TGAT: both the GPU utilization and memory usage increase when sampling more neighborhood nodes. (b) TGAT: the GPU utilization keeps stable and memory usage increases when increasing the mini-batch size. (c) TGN: GPU utilization decreases and memory usage increases when using a larger batch size. (d) MolDGNN: GPU utilization keeps stable and memory usage increases when increasing batch size.**

## 4.1. Temporal data dependency bottleneck

**4.1.1. Bottleneck analysis.** In DGNNs, temporal information is the cause of temporal data dependencies between different modules because the computation must adhere to the sequence of time. In a discrete-time dynamic graph neural network, all node and edge embeddings are processed at once for a single snapshot. Only after all of the node and edge embeddings have been updated will the next snapshot start to be processed. In continuous-time dynamic graph neural networks, updates of the node and edge embedding should occur serially following the time of the events. Thus, temporal data dependency updates embeddings with a restricted order to preserve the meaning of the temporal information. This sequential dependency, unfortunately, cannot be executed in parallel, which limits the computing capability of GPU.

**4.1.2. Experiment Results.** EvolveGCN & MolDGNN both belong to discrete time DGNNs. Their temporal data dependencies exist between GNN and RNN. In EvolveGCN, the temporal data dependency is between RNN and GNN when processing one snapshot at a certain time step because GNN needs to wait for the updated weights from RNN. Likewise, GNN and RNN in the following time steps also need to wait for the results from the previous time steps. As a result, GNN and RNN for processing the same snapshot inside a certain time step and in different time steps cannot be executed in parallel. Similar to EvolveGCN, the RNN in MolDGNN needs to be stalled until the node embedding has been prepared by GNN in the same time step. In addition, since RNN is executed in a time-sequential manner, only one snapshot is processed on GPU at a time. This data dependency between GNN and RNN will cause idle time in GPU computation units. Our profiling results show that during the inference, the GPU utilization of both models is less than 1%, which means the computing resources of GPU are not fully utilized. More details about GPU utilization of MolDGNN are shown in Fig. 6 (d).

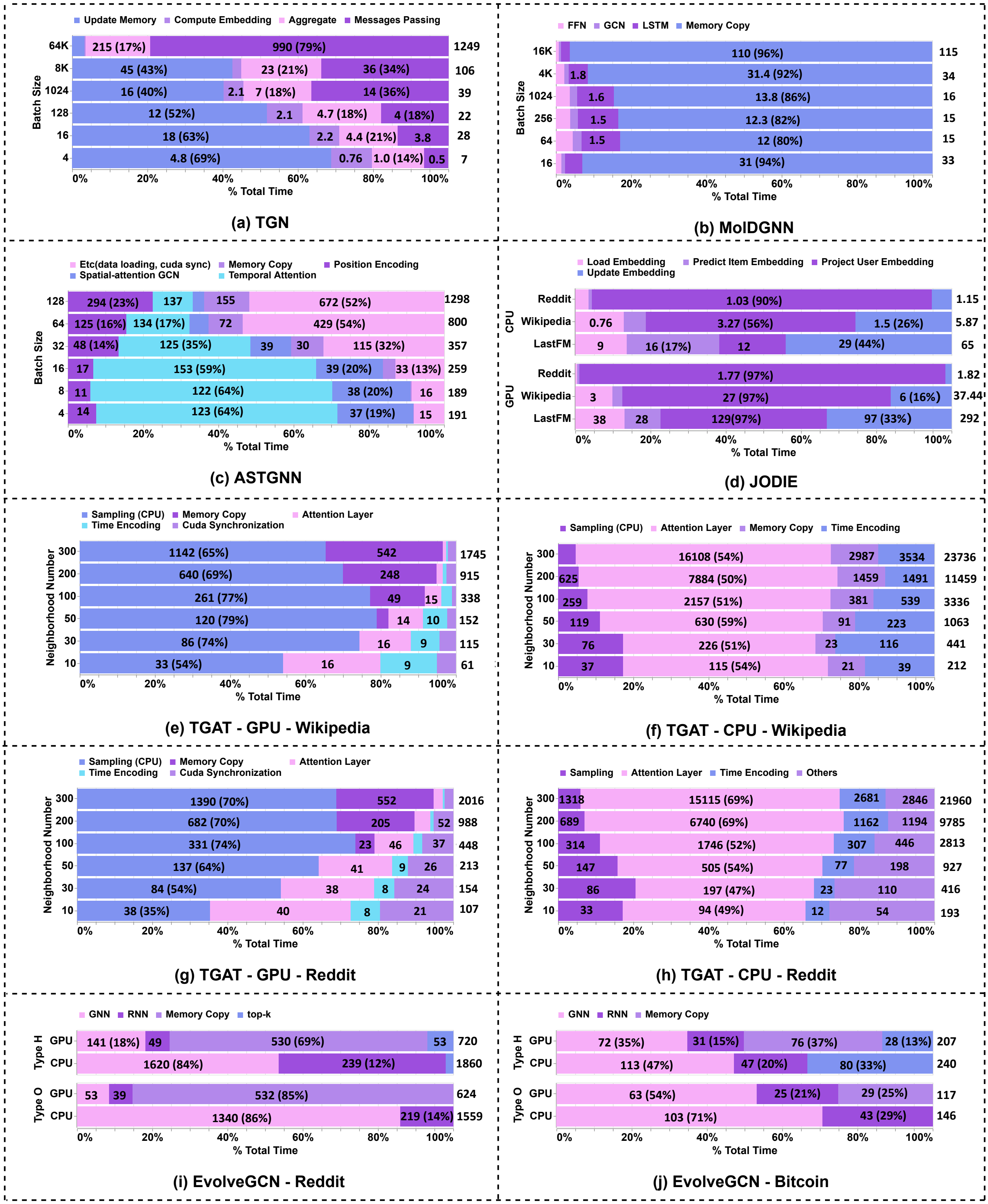

Figure 7: The inference breakdown of different DGNNs in one iteration. We have annotated the execution time (ms) and the proportion of different modules to the whole inference time.

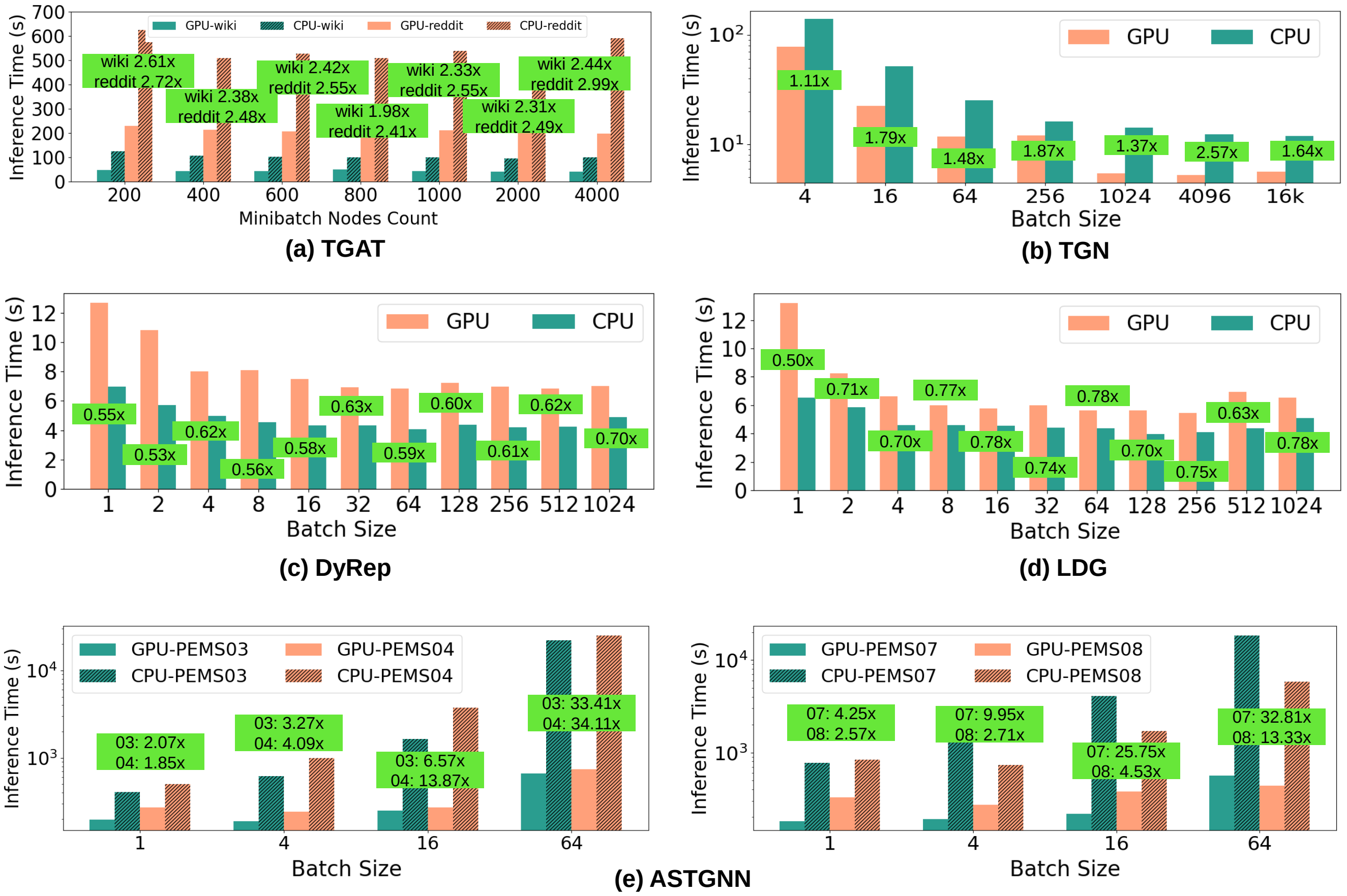

**Figure 8: The inference time comparison and GPU speedup against CPU of different DGNNs.**

**JODIE & TGAT** are both continuous time DGNNs, but the type of their temporal data dependency is slightly different. In TGAT, when updating a node embedding, since the neighborhood node information and temporal encoding information are the prerequisites of the attention layer, the attention layer is always executed after these two modules in a serial manner. Also, TGAT needs to traverse the nodes in one mini-batch in sequential order according to the time the events happen. This temporal data dependency accounts for the low GPU utilization of TGAT (5% $\sim$ 6%) during the inference process, as shown in Fig. 6 (a) and (b). In JODIE, since the output of the previous time RNNs is to be used as the input to the RNNs for the next time step, RNNs in the following steps need to be stalled until all RNNs in the previous steps finish the computation. The data dependency between RNNs makes the GPU utilization of JODIE also very low (around 1.5% $\sim$ 2.5%), even using the t-batch to help increase the parallelism.

**DyRep & LDG** are using the same basic structure for the inference, as shown in Fig. 4 (a) and (b). As shown in Fig. 8 (c) and (d), running inferences on GPU does not outperform the CPU in all batch sizes. Moreover, the profiling experiment reveals the low utilization of GPU. The average utilization is lower than 2% for DyRep, LDG with MLP encoder, and LDG with bilinear.

The major bottleneck comes from the evolving property of the DyRep and LDG. Computing the conditional intensity function at a certain time requires the most recently updated node embedding. Each prediction needs to wait for the previous one to be finished. Two major phases, updating node embedding and computing conditional intensity, are executed in sequence for different time steps. This property disables parallel computing on GPU.

## 4.2. Workload imbalance bottleneck

**4.2.1. Bottleneck Analysis.** The workload imbalance between CPU and GPU is another reason that degrades the hardware performance of DGNNs. It usually comes from two aspects. The first is that some DGNNs behave differently when using different batch sizes. When using a small batch size, the GPU CUDA kernel cannot be saturated with data, which makes most of the kernels idle. When using a large batch size, there will be intensive kernel computation and the CPU will be idle.

The second aspect that greatly affects the hardware performance of DGNNs is the workload imbalance caused by graph sampling. Since the sampling in DGNN also needs to take the temporal information into consideration, it sometimes requires sorting operations and is usually more complex than sampling in static GNNs. In addition, the complexity of the temporal sampling algorithm will cause irregular memory accesses, which will greatly slow down the computation. Usually, DGNNs sample on the CPU and then transfer the data to GPU after the sampling. In this situation, the CPU is occupied by the sampling process. GPU is waiting for the data transmission from the CPU so that it cannot be fully utilized, which causes workload imbalance.

**4.2.2. Experiment Results. EvolveGCN & TGAT** have significant performance degradation with sampling. In EvolveGCN, because of the dimension inconsistency between the node embedding and weight matrix, $k$ node embedding needed to be sampled from all of the node embeddings in the matrix to match the dimension of the weight matrix. To implement this, EvolveGCN uses a fully-connected layer, which we refer to as "top-k", which consumes a considerable execution time with GNN computation, as shown in Fig. 7 (i) and (j).

In TGAT, since a fixed number of neighborhood nodes are needed to be sampled for message aggregation following the temporal sequence, a bisection method and a node index sorting are used, which cause irregular memory accesses and thus lead to a huge computation overhead.

Fig. 7 (e) and (g) present the execution time breakdown of different parts in TGAT with different numbers of sampled neighborhood nodes. Regardless of the number of sampled neighborhood nodes, the neighborhood sampling algorithm on CPU always takes most of the inference time.

The inference time is studied under different mini-batch sizes. We find that neighborhood sampling also highly exceeds other parts in terms of execution time. When the number of nodes in the mini batched changed from 200 to 4,000, the proportion to the total inference time increased from 83% to 94%. Since sampling dominates the inference process, GPU needs to wait for the embedding after sampling from CPU. This has two effects. First, when we increase the mini-batch size, total inference time on the whole dataset does not drop as would seem intuitive, but always remains nearly the same, which is shown in Fig. 8 (a). This is because the data flow is congested in neighborhood sampling on CPU. Second, one is that the GPU utilization keeps constant when increasing the mini-batch size, as shown in Fig. 6 (b). This is further evidence that the sampling on CPU hampers the performance boost of GPU.

**ASTGNN**'s total execution time for the temporal attention component is more than three times that of the spatial GCN component. This is illustrated in the breakdown of the model's inference in Fig. 7 (c). For small batch sizes, temporal attention dominates the entire computation, while CUDA synchronization causes significant delays for larger batch sizes. As shown in Fig. 9, smaller batches cause inefficiencies, making GPU idle for the earlier prediction step. In comparison, larger batches lead to data congestion in PCIe, and the decoder will take a longer time to finish.

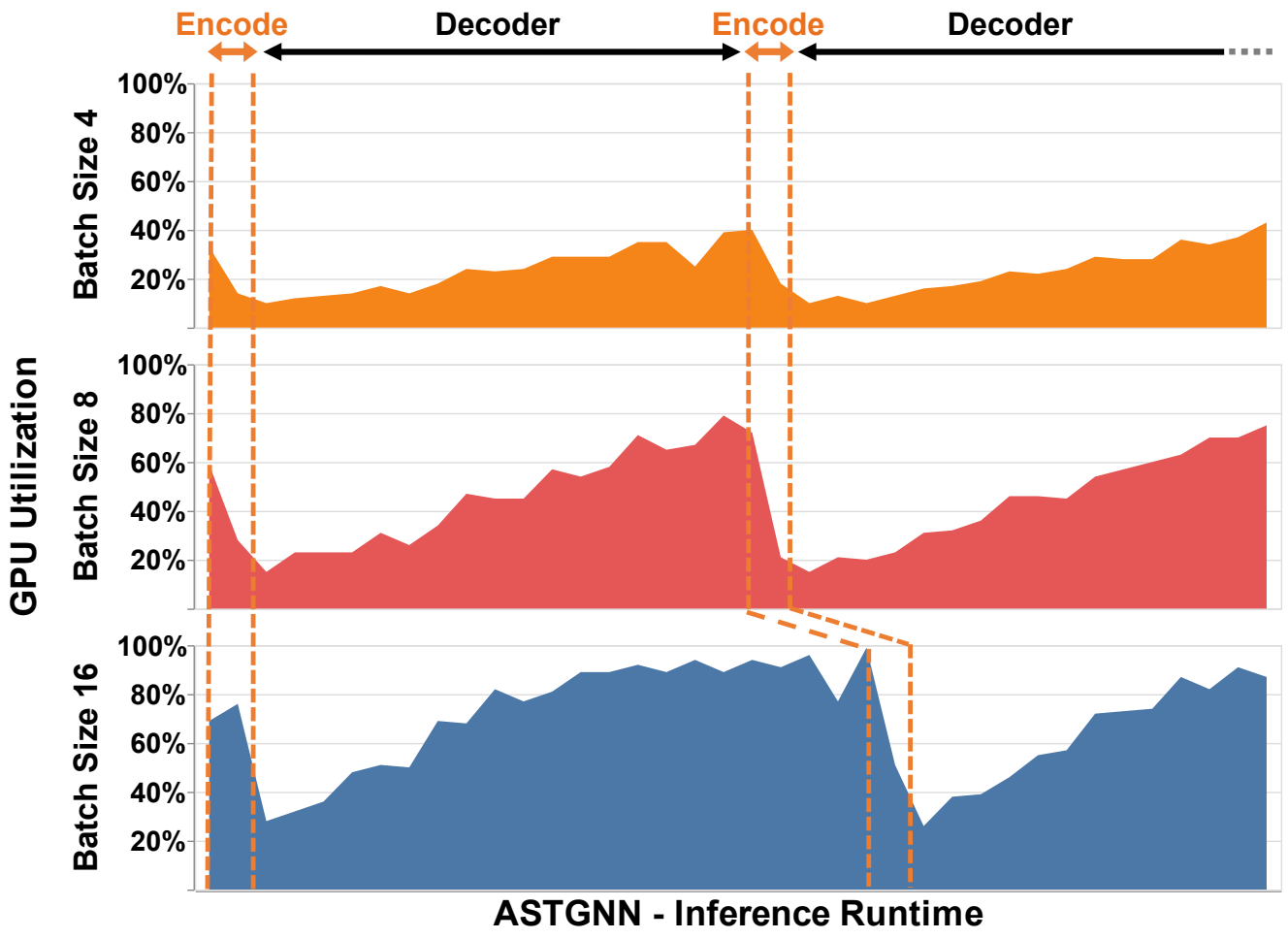

**Figure 9: GPU utilization of ASTGNN inference with batch size (number of subgraphs to be processed at the same time) (a) 4 (b) 8 (c) 16 for two iterations. When the batch size is 16, the encoding stage in the second iteration is delayed because GPU is nearly fully utilized and CPU needs to wait for the completion of GPU computation.**

## 4.3. Data Movement Bottleneck

**4.3.1. Bottleneck analysis.** Data movement between CPU and GPU is another significant bottleneck in dynamic graph neural networks, causing a performance degradation.

In discrete time DGNNs, the cause of the large data movement overhead is the dynamic feature of both the graph topology and node embedding. Every time when computing on a new snapshot, first CPU sends the information of current snapshot from to GPU and then fetch the information from the next time step's snapshot. However, when computing static GNNs on GPU, the graph topology and embedding information only need to be transferred to GPU once. This frequent data exchange can be problematic because the CUDA kernel needs to wait until all required data are on-chip. In continuous time DGNNs, when the mini-batch size is large, the data movement overhead cannot be neglected as well.

**4.3.2. Experiment results. TGN**'s message passing requires frequent data communication between CPU and GPU (Fig. 5 (b)). As shown in Fig. 7 (a), message passing dominates the inference time (79%) when using a larger batch size (64k). This is because when using a larger batch size, more data is needed to be transferred between CPU and GPU , which consumes a lot of GPU working time. This is because not all nodes needed during message passing are in GPU memory. This explains why the GPU utilization of TGN is decreasing when increasing the batch size shown in Fig. 6 (c).

**TGAT** also shows an obvious data movement bottleneck when the sampled neighborhood node count is relatively large. As presented in the model breakdown analysis in Fig. 7 (e) and (g), when the sampled neighborhood nodes count exceeds 100, time spent on data movement will increase rapidly. This is because sampling is executed on the CPU and the computation of the time encoder and attention layer is on GPU, and the larger size of sampled neighborhood nodes lead to a huge amount of data transferring between CPU and GPU.

**EvolveGCN** uses more memory when inferring on larger graphs. This is due to increased memory usage for copying data during inference. As shown in Fig. 7 (i) and (j), the proportion of memory copy using the Reddit dataset is much higher than using the Wikipedia dataset. The reason is that EvolveCGN chooses to reload the snapshots in different time steps from CPU to GPU rather than renew the information on-chip, and the average graph size in the Reddit dataset is larger than that in the Wikipedia dataset. This frequent data exchange causes a large data communication overhead, especially when using large graphs.

**MolDGNN** model does not run efficiently on GPUs. It spends the majority of time on data movement rather than computation. The bottleneck would mainly be the transmission of large number of adjacency matrices of each graph snapshot from CPU to GPU for computation. Once computation is done, the sequence of predicted adjacency matrices will be sent back to CPU for atom-to-atom distances calculation, allowing for chemical simulation. As shown in Fig. 7 (b), whichever batch size we choose, the memory copy occupies about 80% $\sim$ 90% of the GPU working time. This inefficiency occurs when transferring large data between the CPU and GPU, as the devices can end up waiting for data to be transferred between them.

## 4.4. GPU warm-up bottleneck

**4.4.1. Bottleneck analysis.** Although CUDA kernels can greatly accelerate the matrix multiplications thanks to their parallel computing capability, GPU itself can also bring some additional overheads compared to pure inference on CPU. We call these additional overheads GPU warm-up overheads.

The GPU warm-up overheads can be divided into two parts. One is **model initialization** before the inference begins. During this period, GPU will initialize and allocate enough memory on-chip and load all of the weights of the model from CPU memory to GPU via PCIe. During this process, the CUDA runtime API performs stream capture, i.e., building a data flow graph of the network. The whole process can be explicitly done by calling the *cudaStreamCapturing*() function.

Another is GPU **lazy initialization** before the first iteration of inference. CUDA runtime software library creates a context through deferred initialization. When calling each CUDA library function, it will check whether there is a current context; if a context is needed, it will be created automatically. The CUDA runtime API can also force an explicit initialization of the context by calling *cudaFree*() function.

**TABLE 2: GPU warm-up overhead of TGN and MolDGNN. The time of warm-up (ms) and its proportion to the GPU total working time is shown in this table.**

| batch size | TGN | | MolDGNN | |
|---|---|---|---|---|
| | warm-up | computation | warm-up | computation |
| 8 | 5.5 (1%) | 593.2 (99%) | 5.5 (5%) | 94.4 (95%) |
| 32 | 5.3 (3%) | 186.3 (97%) | 10.2 (29%) | 24.6 (71%) |
| 128 | 5.6 (7%) | 80.3 (93%) | 9.8 (55%) | 7.9 (45%) |
| 512 | 5.4 (19%) | 23.7 (81%) | 10.3 (84%) | 2.0 (16%) |
| 2048 | 5.7 (22%) | 20.5 (78%) | 9.8 (93%) | 0.7 (7%) |
| 8192 | 5.5 (48%) | 6.0 (52%) | 9.8 (88%) | 1.3 (12%) |

**4.4.2. Experiment Results.** All of our eight DGNNs have GPU warm-up overhead from our profiling results analysis, and we take TGAT, EvolveGCN, TGN, and MolDGNN as examples for illustration. We have two findings. EvolveGCN and TGAT profiling results show that the GPU warm-up is a one-time overhead. This overhead occurs before the inference starts and the time is constant regardless of the dataset size. Although the proportion of GPU warm-up decreases when the amount of data used for inference gets larger, the GPU warm-up takes about 6.8s, 6.9s, and 6.6s, which is 86×, 41×, and 33× the inference time of processing one mini-batch in TGAT, and one snapshot in EvolveGCN-O and EvolveGCN-H, respectively. Besides, the model initialization on GPU takes about 40×, 855×, and 937× compared to CPU in TGAT, EvolveGCN-O, and EvolveGCN-H, respectively. The second is that from TGN and MolDGNN profiling results, the GPU warm-up overhead proportion to the GPU total working time is increasing with a larger batch size. This is because the GPU will take a longer time to allocate enough memory to store more data on-chip before computation starts. More details are shown in Table. 2.

# 5. Potential Optimizations

According to the hardware bottlenecks of DGNNs that we have discovered, we propose some possible optimization schemes from three aspects: software, hardware, and software-hardware co-design to improve DGNN hardware performance.

A limitation of this work is that we have not yet verified the optimizations suggested in this section through experiments, which are expected to be experimented in the near future. We believe that these optimizations would be highly effective in eliminating DGNN bottlenecks on hardware.

## 5.1. Software optimizations

**5.1.1. Reduce graph preprocessing overhead.** Some dynamic graph preprocessing procedures, such as graph partition, neighborhood node sampling, and time encoding overhead, can to be optimized by executing them in parallel or overlapping them with other modules. For example, Zhang et al. [39] hide the sampling overhead by overlapping the sampling and GNN inference on the CPU-FPGA heterogeneous platform. Jangda et al. [40] accelerate graph sampling on GPU using multi-threads. Another way to reduce graph preprocessing overhead is to improve the memory locality. This is challenging because dynamic graph topology will cause irregular memory access. Although GNNSampler [41] avoids the long memory access time by sampling nodes that have almost contiguous addresses with a higher probability to increase the memory locality, it can only be used in static GNNs, and there is still a large room for DGNNs memory locality improvement.

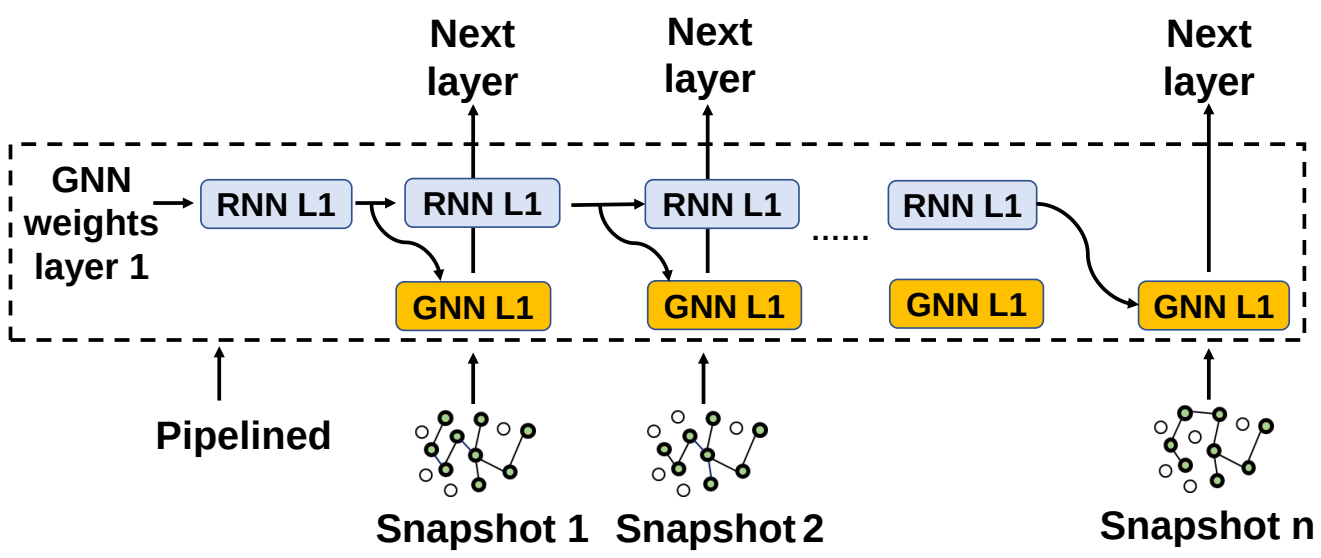

**Figure 10: The RNN and GNN in adjacent time steps are pipelined in EvolveGCN.**

**5.1.2. Smarter PyTorch-like software framework.** In order to take advantage of the parallelism offered by GPUs, software must be written in a way that is aware of the underlying hardware. This can be a challenge for developers, as it requires a deep understanding of the architecture and how to optimize for it. The ideal solution is to use a software framework that is both hardware-aware and software-aware. Such a framework can automatically optimize for data movement, pipelining, and other factors based on the particular GPU and the workload being run. This can greatly reduce the complexity of writing GPU code and can result in significantly better performance.

## 5.2. Hardware optimizations

**5.2.1. Improve parallelism.** Although some modules in DGNN, such as RNN, GNN, and time encoding, have temporal data dependencies within the same time step, the parallelism of DGNN execution on hardware can be improved by paralleling modules of different time steps. For instance, we can parallel the GNN and RNN belonging to different time steps in EvolveGCN, as suggested in Fig. 10. In addition, the neighborhood sampling and attention layer in TGAT, the updating phase and the computing phase in LDG can also be pipelined similar to EvolveGCN.

**5.2.2. Reduce data transferring overhead.** In DGNNs, the changing graph topology will lead to node and edge embeddings varying with time. As a result, we cannot load all of the node and edge information at every time step to the on-chip memory once because of huge amount of data. From our observation in the profiling experiment, DGNNs usually have a frequent memory exchange between CPU and GPU. A possible solution to reduce the data movement overhead is to find the similarity between different snapshots between adjacent time steps. This avoids duplicate data communication by transferring only part of the node and edge information which have been updated. For example, EvolveGCN uses sliding time windows [42] in the graph prepossessing process to overlap snapshots. The increased similarity by overlapping will lead to a less radical change from one snapshot to the next, which makes the solution possible. Therefore, reducing data transferring for DGNNs hardware implementation is a promising direction for future research.

## 5.3. Hardware/Software co-design

**5.3.1. Extend static GNN hardware optimizations to DGNNs.** Since dynamic graph structures in some DGNNs can be converted into static ones by applying graph preprocessing before inference starts, it is possible for future dynamic GNN accelerators to make some modifications to current static GNN accelerator optimizations. For example, when designing an accelerator for discrete time DGNNs, graph convolution extracts spatial information from each snapshot, and there already exists mature optimizations for graph convolution on static graphs. Therefore, a quick optimization strategy is to combine the current static graph optimizations with new proposed optimizations for temporal information processing to achieve a better hardware performance.

**5.3.2. Design hardware-efficient DGNNs.** Current DGNNs are designed to improve software performance and are not hardware-friendly. Therefore, it is crucial to design hardware-efficient DGNNs specifically for hardware performance improvement. While little has been published on hardware optimizations for DGNNs, two articles [43, 44] illustrate model compression methods, which can be extended to dynamic graphs. These two works are a good start for model-level dynamic graph acceleration.

# 6. Conclusions and Future Work

In this paper, we conduct a comprehensive survey of the current state of DGNN development and point out that the hardware acceleration and optimization methods for static GNN proposed by previous researchers are not suitable for DGNN. We select eight DGNNs with different characteristics and profile them on CPU and GPU, respectively. Based on the experimental results, we summarize four common hardware bottlenecks of DGNNs: temporal data dependencies, workload imbalance, data movement and GPU warm-up. We also propose several potential optimizations that are possible to alleviate the DGNN hardware bottlenecks from three aspects: software, hardware, and software/hardware co-design, respectively.

Although this work is a comprehensive review of the performance of DGNNs on hardware, we only consider the hardware bottlenecks during inference. It is likely that new hardware bottlenecks will be found during training, which is more complex than inference. Therefore, our future research will focus on profiling the training of DGNNs. In addition, we plan to design a DGNN hardware accelerator to realize the optimizations proposed in this paper and push the boundaries of efficient DGNN hardware design.

# A. Artifact Appendix

## A.1 Abstract

This artifact appendix provides guidelines to obtaining the dynamic graph neural network profiling results on CPUs and GPUs. We elaborate on the software environment and the target hardware devices used in our experiment. We also provide the source code used in our experiments with examples for the generated profiling results.

## A.2 Artifact check-list (meta-information)

- **Algorithm:** Dynamic graph neural network inference profiling.
- **Program:** Pytorch Profiler and NVIDIA Nsight Systems.
- **Compilation:** PyTorch.
- **Model:** EvolveGCN, TGAT, JODIE, TGN, MolDGNN, AST-GNN, DyRep, LDG.
- **Data set:** Bitcoin Alpha, Reddit Hyperlink Network, Stochastic Block Model, Wikipedia, Caltrans Performance Measurement System (PeMS), Social Evolution, GitHub-data, LastFM, and Molecular Trajectories ISO17 Dataset.
- **Run-time environment:** Red Hat Expertise Linux 4.18.0-372.9.1.el8.x86_64
- **Hardware:** CPU: Intel(R) Xeon(R) Gold 6226R CPU @ 2.90GHZ. GPU: NVIDIA RTX A6000.
- **Run-time state:** : Run-time varies based on model type and configuration.
- **Execution:** Executing commands can be found on GitHub.
- **Output:** PyTorch Profiler and NVIDIA Nsight Systems trace files.
- **Experiments:** Analyzing the profiling results of the inference phase of each model and investigating GPU behavior and data-dependency of the models.
- **How much disk space required (approximately):** 80GB
- **How much time is needed to prepare workflow (approximately):** About one hour to install related python packages and NVIDIA Nsight Systems.
- **How much time is needed to complete experiments (approximately):** About 2 hours to profile one model on both CPU and GPU with different parameters.
- **Publicly available:** Yes.
- **Code licenses (if publicly available):** Creative Commons Attribution 4.0 International
- **Workflow framework used:** PyTorch.
- **Archived (provide DOI):** 10.5281/zenodo.7079621

## A.3 Description

### A.3.1 How to access

Our source code and profiling results are available at Zenodo (`https://zenodo.org/record/7079621#.YyKMznbML3H`) and GitHub (`https://github.com/eun4231/DGNN_analysis`)

### A.3.2 Hardware dependencies

We recommend testing on CPU (Intel(R) Xeon(R) Gold 6226R CPU @ 2.90GHZ) and GPU (NVIDIA RTX A6000). Any similar hardware platforms will provide similar results.

### A.3.3 Software dependencies

The operation system we used is Red Hat Expertise Linux 4.18.0-372.9.1.el8.x86_64. The software framework is PyTorch. The implementation of each model is from publicly available source code from original authors while we made some modifications for easy hardware profiling. The versions of software and hardware drivers in the experiment are shown below:

```
conda == 4.11.0
Python == 3.9.7
requests == 2.26.0
rhel == 7.9
glibc == 2.17
cudnn = 11.7
```

The profiling and analysis of profiling results is based on two hardware profiling tools: PyTorch Profiler (`https://pytorch.org/tutorials/recipes/recipes/profiler_recipe.html`) and NVIDIA Nsight Systems (`https://developer.nvidia.com/nsight-systems`). Pytorch Profiler is a simple profiler API integrated in the PyTorch framework. Tensorboard needs to be installed to view the generated trace file from PyTorch Profiler. NVIDIA Nsight Systems Workstation needs to be installed to do the profiling with it and view the trace file from this system.

The version of the Python packages we used is shown as follows:

```
matplotlib==3.5.2
numpy==1.21.2
pandas==1.4.2
PyYAML==6.0
scikit_learn==1.1.2
scipy==1.8.1
torch==1.10.1
torchvision==0.11.2
tqdm==4.64.1
```

### A.3.4 Data sets

Bitcoin Alpha
(`http://snap.stanford.edu/data/soc-sign-bitcoin-alpha.html`)
Reddit Hyperlink Network
(`http://snap.stanford.edu/data/soc-RedditHyperlinks.html`)
Stochastic Block Model
(`https://github.com/IBM/EvolveGCN/tree/master/data`)
Wikipedia
(`https://snap.stanford.edu/jodie/`)
Caltrans Performance Measurement System (PeMS)
(`https://dot.ca.gov/programs/traffic-operations/`

mpr/pems-source)
LastFM (http://snap.stanford.edu/jodie/lastfm.csv)
Molecular Trajectories ISO17 Dataset (http://quantum-machine.org/datasets)
Social Evolution (http://realitycommons.media.mit.edu/socialevolution4.html)
GitHub-data (https://www.gharchive.org/)

#### A.3.5 Models

EvolveGCN, TGAT, JODIE, TGN, MolDGNN, ASTGNN, DyRep, LDG

### A.4 Installation

To install PyTorch Profiler and Tensorboard, run this command:

```
pip install torch torchvision
```

For more information related to PyTorch Profiler, please refer to the tutorial (https://pytorch.org/tutorials/recipes/recipes/profiler_recipe.html).

To install NVIDIA Nsight Systems, download the installer from this link: https://developer.nvidia.com/gameworksdownload#?dn=nsight-systems-2022-3 and run the installer.

### A.5 Experiment workflow

Once the installation is completed, run the following commands to do profiling with PyTorch Profiler. Below is an example of profiling the EvolveGCN model; the parameters in the commands can be changed to get different profiling results. For more elaborated examples of how to profile the other models, refer to the code on GitHub.

```
# to run EvolveGCN-O with PyTorch Profiler
python run_exp.py --config_file ./experiments/
parameters_bitcoin_alpha_edgecls_egcn_o.yaml

#to run EvolveGCN-H with PyTorch Profiler
python run_exp.py --config_file ./experiments/
parameters_bitcoin_alpha_edgecls_egcn_h.yaml

#to run TGAT with PyTorch Profiler
python -u learn_edge.py -d wikipedia --bs 200
--uniform  --n_degree 20 --agg_method attn
--attn_mode prod --gpu 0 --n_head 2
--prefix hello_world --use_cuda True
--start_profiling True

#to run TGAT with NVIDIA Nsight Systems
nsys profile -w true -t cuda,nvtx,osrt,cudnn,
cublas -s cpu --capture-range=cudaProfilerApi
--capture-range-end=stop
--cudabacktrace=true -x true
--cuda-memory-usage=true
python -u learn_edge.py -d wikipedia --bs 200
--uniform  --n_degree 20 --agg_method attn
--attn_mode prod --gpu 0
--n_head 2 --prefix hello_world
```

### A.6 Evaluation and expected results

After correctly running the commands in the experiment workflow subsection, PyTorch Profiler users will get trace files end with *.pt.trace.json. Run the following command to install PyTorch Profiler TensorBoard Plugin and open the trace file:

```
pip install torch_tb_profiler
tensorboard --logdir=file_directory
```

The file_directory is the path of the trace file. One can expect profiling results as follows:

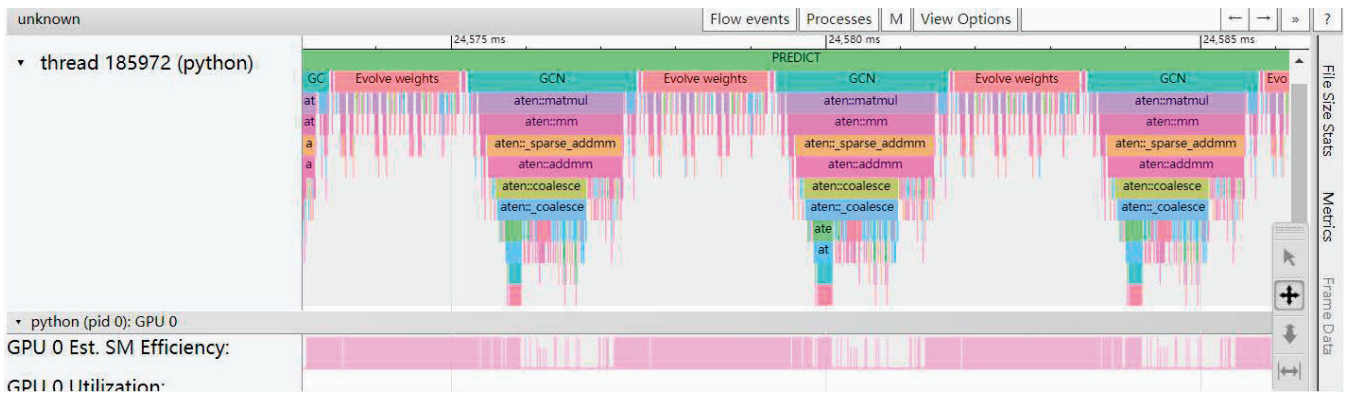

**Figure 1.** EvolveGCN profiling results using Pytorch Profiler.

NVIDIA Nsight Systems users will find trace files ending with *.nsys-rep, which can be directly opened using Nsight Systems. One can expect profiling results as follows:

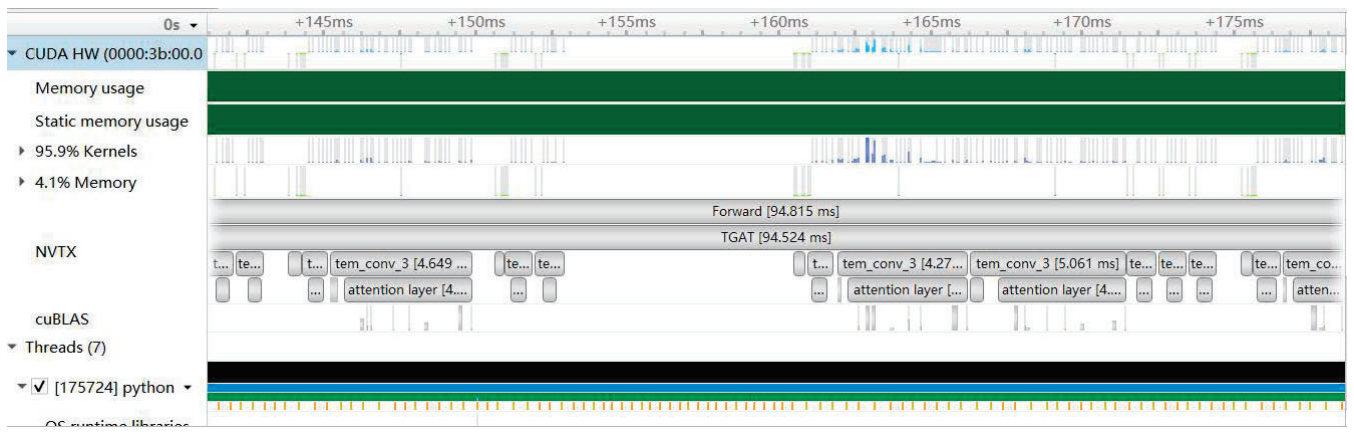

**Figure 2.** TGAT profiling results using NVIDIA Nsight Systems.

### A.7 Methodology

Submission, reviewing and badging methodology:

- https://www.acm.org/publications/policies/artifact-review-badging
- http://cTuning.org/ae/submission-20201122.html
- http://cTuning.org/ae/reviewing-20201122.html